\documentstyle[12pt,epsf]{article}

\begin{document}

\vspace{1.3cm}

\begin{center}
{\LARGE \bf
Luttinger liquid behavior in single-wall nanotubes}
\end{center}

\vspace{0.5cm}

\begin{center}
{\large
Andrei Komnik and Reinhold Egger}

\vspace{0.5cm}
 
{\small \em
Fakult\"at f\"ur Physik, Albert-Ludwigs-Universit\"at, 
 D-79104 Freiburg}
\end{center}

\vspace{0.7cm}

{\small
\begin{quote}
Transport properties of metallic single-wall
nanotubes are examined based on the Luttinger liquid theory.
Focusing on a nanotube transistor setup, the linear conductance 
is computed from the Kubo formula using perturbation
theory in the lead-tube tunnel conductances.  
For sufficiently long nanotubes and high temperature, 
phonon backscattering should lead
to an anomalous temperature dependence of the resistivity.
\end{quote}
}
\vspace{0.4cm}

Carbon nanotubes possess many fascinating properties
and have recently attracted a lot of attention.
Metallic nanotubes can behave as ballistic
quantum wires over lengths of several $\mu$m 
\cite{tans1} and hence constitute perfect 
experimental realizations of 1D conductors.  
It is now well-known that Fermi liquid theory must break down in such
a 1D conductor because of the Coulomb interactions among the 
electrons. In fact, at temperatures above the milli-Kelvin range,
an individual metallic single-wall nanotube (SWNT)
can be accurately described in terms of Luttinger liquid (LL)
theory \cite{egger}.  The LL is a prototypical yet simple
model for non-Fermi liquid behavior.
The interaction strength is measured
in terms of a single parameter $g\leq 1$, where $g=1$ is the noninteracting
limit. For a SWNT, the theoretical estimate is \cite{egger}
$g \simeq \{ 1+ 
(8e^2/\pi\hbar\epsilon v_F) \ln(L/2\pi R) \}^{-1/2}$,
where the only logarithmic dependence on the tube length $L\approx 1~\mu$m
 and the radius $R\approx 1.4$~nm leads to a value
 around $g\approx 0.2$ to $0.3$.
Here, $\epsilon$ is the background dielectric constant
and $v_F\approx 8\times 10^5$~m/s is the Fermi velocity.
The LL should also show up in a variety of other systems, such as
long chain molecules, the edge states in a fractional
quantum Hall bar, or in single-channel quantum wires
in semiconductor heterostructures.  Unfortunately, 
despite of many efforts, a generally accepted experimental realization
of the LL in these systems is still lacking.
For a nanotube rope, however, transport
experiments have recently been reported that provide clear
evidence for a LL \cite{bockrath}.  A rope should show LL behavior
if three conditions are met: (a) only one metallic 
tube is contacted by the leads, (b) most tubes in the
rope are not metallic, and (c) electron tunneling 
between different metallic tubes in the rope can be
neglected. 

Let us consider a transistor consisting of a tube of length $L$,
where contact to external leads is established 
 at positions $0<x_1<x_2<L$.  The leads $i=1,2$ are 
modelled by free electrons, and 
we take the standard tunneling Hamiltonian for the lead-tube couplings.
We assume pointlike contacts and focus on the linear dc conductance,
which, according to the Kubo formula, takes the form
\begin{equation}\label{kubo}
G = \lim_{\omega\to 0} \frac{1}{\hbar \omega}  
{\rm Im}\left[
\int_0^{\hbar \beta} d\tau\, \exp(i\Omega\tau)\,
 \langle I^{(1)}(\tau) I^{(2)}(0)  \rangle
 \right]_{i\Omega\to \omega+i\delta} \;,
\end{equation}
where $\beta=1/k_B T$. 
Since the dc current through both contacts coincides,
we may take an arbitrary linear combination 
$I^{(1)}=\epsilon_1 I_1 + \epsilon_2 I_2$  [where $\epsilon_1+
\epsilon_2=1$] of the
currents $I_1$ and $I_2$ through the tunnel contacts
at $x_1$ and $x_2$, respectively.
Since the transport voltage can also be split up arbitrarily \cite{georg},
the second current operator can be written as
 $I^{(2)}=\kappa_1 I_1+\kappa_2 I_2$
with $\kappa_1+\kappa_2=1$.  This gives 
\begin{equation} \label{cond}
G = \sum_{ij=1,2} \epsilon_i \kappa_j G_{ij} \;,
\end{equation}
where the matrix elements $G_{ij}$ directly follow from Eq.~(\ref{kubo}).
We compute these matrix elements by perturbation theory in
the dimensionless bare tunnel conductances $g_i=R_K/R_{T,i}$
 with the resistance quantum  $R_K=h/e^2$ and 
the tunnel resistance $R_{T,i}$ through contact $i=1,2$.
 Under an exact calculation,
one could use any choice for $\epsilon_1$ and $\kappa_1$
in Eq.~(\ref{cond}) \cite{georg}.
Under a perturbative calculation for the $G_{ij}$,
however, requiring independence
of $\epsilon_1$ and $\kappa_1$ is equivalent to maximizing 
Eq.~(\ref{cond}) with respect to these parameters and gives
$\epsilon_1=(G_{21}+G_{22})/\sum_{ij} G_{ij}$ and
$\kappa_1=(G_{12}+G_{22})/\sum_{ij} G_{ij}$.
The lowest-order result in $g_i$  follows from the
expansion $G_{ii}=G_i+\delta G_{ii}$, where $\delta G_{ii}$
as well as $G_{12}$ and $G_{21}$ are of at least second order
in $(g_1,g_2)$ but $G_i$ is of first order.
Then Eq.~(\ref{cond}) leads to  
\begin{equation}  \label{seq}
G = \frac{G_1 G_2}{G_1+G_2} \;.
\end{equation}
This formula describes sequential incoherent transport
through the device and is appropriate for high temperatures,
where $k_B T$ exceeds the charging energy
 $E_c\approx e^2 \ln(L/R)/\epsilon L$.
In addition, the condition
$\hbar v_F/k_B T \ll |x_2-x_1|$ should hold.   
On the other hand, at lower temperatures coherent processes
such as co-tunneling play a prominent role
and lead to the breakdown of Eq.~(\ref{seq}). 
From the general expression (\ref{cond}) we observe that
$G\approx G_{12}/2$, since $G_i$ vanishes as $T\to 0$
unless one has a resonance.
This is consistent with the results 
of Ref.~\cite{kinaret} for transport through a LL ring but at the same time
indicates that the latter are only valid for 
thermal energies $k_B T$ well below the 
charging energy $E_c$.  Therefore the ``straightforward'' application of the
Kubo formula to such problems is not as simple as commonly thought.

\begin{figure}
\hfil
\epsfxsize=8cm
\epsffile{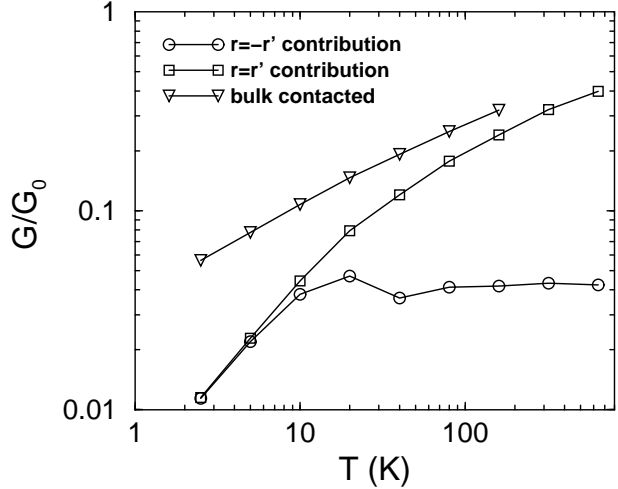}
\hfil
\caption[fig2]{
High-temperature conductance (\ref{seq})
for $g=0.2$, $g_1=g_2=0.1$  and
$L=10^4 a=2.46~\mu$m. Note the double-logarithmic scales.
For the bulk- (end-) contacted case, $x_1=L-x_2=3500 a$
($5 a$).  For the end-contacted case, the $r=r'$ and $r=-r'$ contributions
are plotted separately.  
}
\label{fig2}
\end{figure} 

Let us start with the high-temperature limit.
Perturbation theory yields with $G_0=4e^2/h$ 
and the Fourier transformed LL Greens function $K(x_i,\xi)$ at equal sites,
\begin{equation} \label{final}
G_i/G_0 = -\pi g_i \int d\xi \,( -dn_F(\xi)/d\xi )  
\;{\rm Im}\,K(x_i,\xi) \;.
\end{equation}
The derivative of the Fermi function is $-dn_F(\xi)/d\xi=
(4k_B T)^{-1} \cosh^{-2}[\xi/2k_B T]$.
Technically speaking, $K(x_i,\xi)$ is obtained from a 
decomposition of the electron operator
into 1D fermions $\psi_{r,\alpha,\sigma}$, where $r=\pm$ 
is the right- or left-moving part, $\alpha=\pm$ denotes the
right or left Fermi point, and $\sigma=\pm$ is the spin index,
and subsequent use of the bosonization method \cite{egger}.
Eq.~(\ref{final}) shows that the conductance is related
to the local tunneling density of states of the LL.
The respective power laws are well-known \cite{egger},
$G_i \propto T^\eta$, where the exponent $\eta$ is given
by $\eta_b=(g^{-1}+g-2)/8$ for tunneling into the bulk, i.e.,
far away from the ends of the tube, and the end exponent
is $\eta_e= (g^{-1}-1)/4$, see Figure \ref{fig2}.
These predictions have been
verified in recent experiments \cite{bockrath}.
For the bulk-contacted
case, the slope coincides with the correct value $\eta_b=0.4$
for $g=0.2$.  For the end-contacted case,
the slope is close to the end value $\eta_e=1.0$ for
low temperatures, but exhibits a crossover to the smaller
bulk exponent around $T\approx 50$~K.
This can be rationalized by separately looking at the
$r=r'$ and $r=-r'$ contributions to the conductance.
Mixing of right- and left-movers ($r=-r'$) violates momentum conservation and 
is only allowed close to the end.  Due to thermal decorrelation
such processes are destroyed with increasing temperature.

\begin{figure}
\hfil
\epsfxsize=8cm
\epsffile{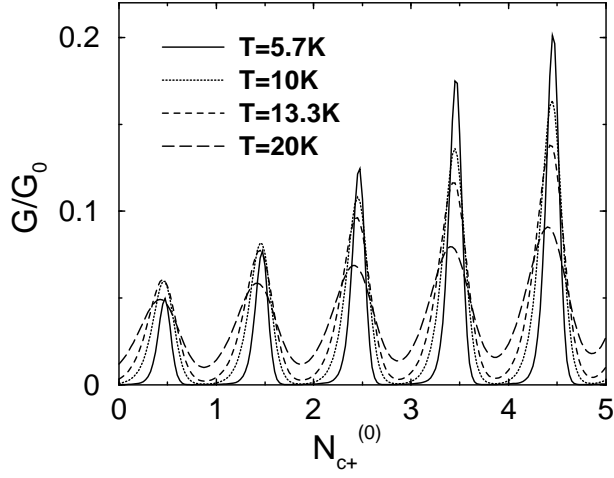}
\hfil
\caption[fig3]{
Low-temperature conductance $G= G_{12}/2$ as a function 
of the gate voltage $V_G\propto N_{c+}^{(0)}$ for
$g=0.2, g_1=g_2=0.1, L=10^4 a$ and $x_1=L-x_2=0.1 L$.
}
\label{fig3} 
\end{figure}

For $k_B T\ll E_c$, we compute the conductance as $G_{12}/2$.
Following Ref.~\cite{kinaret}, we employ the Wick theorem.
 Although strictly speaking
this is not correct, at sufficiently low temperatures  
the corresponding errors in $G$ are expected to become
very small.  The gate-voltage dependence of the conductance
is shown in Fig.~\ref{fig3}.  Due to the Coulomb blockade,
one finds characteristic peaks.  Their lineshape accurately
follows the standard $\cosh^{-2}(\Delta V_G/2k_B T)$ form,
and also the temperature dependence of the peak heights
is in agreement with conventional Coulomb blockade theory.

In the remainder, we investigate the effect of
{\sl phonon backscattering} on the SWNT conductance. 
In most other systems where LL behavior is thought
to be present phonon backscattering does not play
an important role for transport properties because
of the large momentum transfer $2k_F$.
However, in a SWNT the relevant momentum transfer is
$2 q_F$, where $q_F\ll k_F$ is tuned by
the gate voltage. Thermal
population of phonon modes is then much more significant.
The only low-energy phonons that couple right- and
left-movers are the acoustic torsional modes
with dispersion $\Omega(q) = v_t |q|$, where
$v_t\approx 1.4\times 10^4$~m/s.
Following Ref.~\cite{kane2}, the electron-phonon coupling is
\begin{equation}\label{ep1}
H_{e-p}= \lambda \int dx \vartheta(x) \sum_{\alpha\sigma}
\left(\psi^\dagger_{+,\alpha,\sigma}  \psi^{}_{-, \alpha,\sigma} 
+ {\rm H.c.}\right)\;,
\end{equation}
with $\lambda\simeq n\times 2.92$~eV{\AA}
for a $(n,n)$ tube.  
Using the bosonization method to describe the electronic
degrees of freedom $\psi_{r,\alpha,\sigma}$, we focus on
 the lowest-order ($\propto\lambda^2$) contribution
$\delta G_p(T)$ to $G(T)$.
The phonon field $\vartheta(x)$  
can be expressed in terms of free boson operators $a_q$.
Since $v_t\ll v_F$, we can safely 
neglect the time-dependence of $a_q$, and hence the
phonon averaging simply produces a boson mode occupation
factor $\coth[\hbar\beta\Omega(q)/2]$.  Furthermore, for 
exactly the same reason, this factor can be approximated by
$2/\hbar \beta v_t |q| $, and we obtain
\begin{equation} \label{cond2}
\delta G_p (T) / G_0  =  - L \frac{\gamma_g \lambda^2}{a C_t v_F}
(\pi a k_B T/v_F)^{(1+g)/2}  \;,
\end{equation}
with the twist modulus $C_t\simeq n^3 \times 18$~eV{\AA} \cite{kane2}
and the numerical prefactor
$\gamma_g = 4 \pi^{-2} g^{1+g/2}\, \sin[\pi(3+g)/4] 
\int_0^\infty dz \,z/\sinh^{(3+g)/2}(z)$.
For example, $\gamma_1=2/\pi$ and $\gamma_{0.2}\simeq 0.112$. 
The tube length $L$ appearing in Eq.~(\ref{cond2}) should be 
replaced by $|x_2-x_1|$ if the phonon modes can be
pinned at the locations of the tube-lead contacts. 
Further modifications might be necessary if strong interactions
with the substrate influence the phonon dynamics.

To convert Eq.~(\ref{cond2}) into a resistivity, we 
note that $G\equiv G_0 \tilde{g} =G_0+\delta G_p$ is the 
two-terminal conductance of a nanotube in adiabatic
contact to voltage sources.  The resistivity is
obtained from the four-terminal conductance $\widetilde{G}=G_0 
\tilde{g}/(1-\tilde{g})$,
so that the 1D relation $\widetilde{G}=\sigma/ L$ with
$\rho=1/\sigma$ yields the 1D resistivity due to phonon backscattering
\begin{equation}\label{resist}
\rho = \frac{h}{4e^2} \frac{\gamma_g \lambda^2}{a C_t v_F}
(\pi a k_B T/v_F)^{(1+g)/2}  \;.
\end{equation}
For $g=1$, Eq.~(\ref{resist})  agrees 
with the theory for uncorrelated electrons  
and the corresponding experimental results 
found in many-rope systems and mats \cite{kane2}.
For an individual SWNT, the long-ranged Coulomb interactions among
the electrons change this into an anomalous $\rho\sim T^{(1+g)/2}$
power law.
It is apparent from Eq.~(\ref{cond2}) that the conductance
will be dominated by phonon backscattering at sufficiently high temperatures,
$T>T^*_g(L)$.  Let us therefore estimate $T^*$ employing the
{\sl ad hoc} criterion $|\delta G_p|/G_0=\nu$ with, say, $\nu=0.1$. 
This gives
\begin{equation}\label{tstar}
T^*_g (L) \approx 8000 \left( \frac{3560 \gamma_g}{n} \; L 
\right)^{-2/(1+g)} \;,
\end{equation} 
where $T^*$ is measured in Kelvin and $L$ in $\mu$m.  
Putting $n=10$ and $g=1$, this gives $T^* =180$~K for
a $L=200$~nm tube, and $T^* = 35$~K for a $L=1 \mu$m tube.
For strongly correlated electrons,  the length dependence
is  even more dramatic.  Putting $g=0.2$ and again $n=10$,
for a $L=200$~nm tube we get $T^*=250$~K, while for
a $L=1~\mu$m tube, phonons already begin to dominate above $T^*=17$~K.

We thank  G.~G\"oppert, A.~Gogolin and H.~Grabert for 
discussions, and acknowledge support from the DFG.

\end{document}